\documentclass[twocolumn]{jpsj3} 
\usepackage{graphicx}
\usepackage{bm}

\title{Multiple Antiferromagnetic Spin Fluctuations and Novel Evolution of $T_c$ in Iron-Based Superconductors LaFe(As$_{1-x}$P$_x$)(O$_{1-y}$F$_y$) Revealed by $^{31}$P-NMR Studies}

\author{
Takayoshi Shiota$^{1}$, Hidekazu Mukuda$^{1}$\thanks{E-mail: mukuda@mp.es.osaka-u.ac.jp}, Masahiro Uekubo$^{2}$, Fuko Engetsu$^{1}$, Mitsuharu Yashima$^{1}$, \\Yoshio Kitaoka$^{1}$, Kwing To Lai$^{2}$, Hidetomo Usui$^{2}$, Kazuhiko Kuroki$^{2}$, \\
Shigeki Miyasaka$^{2}$\thanks{E-mail: miyasaka@phys.sci.osaka-u.ac.jp}, and Setsuko Tajima$^{2}$
}

\inst{
$^{1}$Graduate School of Engineering Science, Osaka University, Toyonaka, Osaka 560-8531, Japan \\
$^{2}$Graduate School of Science, Osaka University, Toyonaka, Osaka 560-0043, Japan \\
}

\abst{
We report on $^{31}$P-NMR studies of LaFe(As$_{1-x}$P$_x$)(O$_{1-y}$F$_{y}$) over wide compositions for 0$\le x \le$1 and 0$\le y \le$0.14, which provide clear evidence that antiferromagnetic spin fluctuations (AFMSFs) are one of the indispensable elements for enhancing $T_c$. 
Systematic $^{31}$P-NMR measurements revealed two types of AFMSFs in the temperature evolution, that is, one is the AFMSFs that develop rapidly down to $T_c$ with low-energy characteristics, and the other, with relatively higher energy than the former, develops gradually upon cooling from high temperature. The low-energy AFMSFs in low $y$ (electron doping) over a wide $x$ (pnictogen height suppression) range are associated with  the two orbitals of $d_{xz\!/yz}$, whereas the higher-energy ones for a wide $y$ region around low $x$ originate from the three orbitals of $d_{xy}$ and $d_{xz\!/yz}$. 
We remark that the nonmonotonic variation of $T_c$ as  a function of $x$ and $y$ in  LaFe(As$_{1-x}$P$_x$)(O$_{1-y}$F$_y$) is attributed to these multiple AFMSFs originating from degenerated multiple $3d$ orbitals inherent to Fe-pnictide superconductors.
}

\recdate{******}
\newpage
\begin{document}
\maketitle

\begin{figure}[htbp]
\centering
\includegraphics[width=6cm]{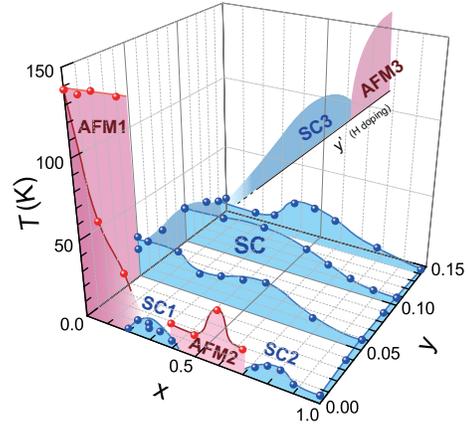}
\caption[]{(Color online) 
Superconducting and antiferromagnetic phase diagram of LaFe(As$_{1-x}$P$_x$)(O$_{1-y}$F$_{y}$) for $y$=0.14 (Uekubo {\it et al.}\cite{Uekubo}) and  0$\le y \le 0.10$\cite{Kamihara2008,Hosono_review,Luetkens,Saijo,Lai_PRB,Wang,SKitagawa_2014,Mukuda_jpsj2014}, 
together with schematic phases of  SC3 and AFM3 for LaFeAs(O$_{1-y'}$H$_{y'}$)~\cite{Iimura_H,Hiraishi}.
}
\label{phasediagram}
\end{figure}

Since the discovery of superconductivity (SC) in a layered iron(Fe)-pnictide LaFeAs(O$_{1-y}$F$_y$)\cite{Kamihara2008}, a number of researches have unraveled a rich variety of antiferromagnetic (AFM), structural, and SC phase diagrams in various Fe-pnictide families\cite{Hosono_review}. 
The SC transition temperature ($T_c$) of LaFe(As$_{1-x}$P$_x$)(O$_{1-y}$F$_y$) exhibits a unique nonmonotonic variation, as shown in Fig. \ref{phasediagram}, where the respective compositions $x$ and $y$ control the local  lattice parameter of the Fe-pnictogen ($Pn$) tetrahedron ($x$) through  the isovalent substitution of As with P  and an electron-doping level ($y$) through the substitution of O$^{2-}$ with F$^{-}$ \cite{Luetkens,Saijo,Lai_PRB,Wang,Uekubo}.  
Previous NMR studies of this series ($0\le y\le0.1$) revealed that the AFM spin fluctuations (AFMSFs)  are markedly enhanced at $x$ where $T_c$ exhibits a peak\cite{Mukuda_PRB2014}.
The appearance of such unexpected AFMSFs was related to a reemergent AFM order phase at ($x$, $y$)=(0.6, 0), denoted as AFM2 in Fig. \ref{phasediagram}, which is separated from the AFM1 at the parent LaFeAsO of ($x$, $y$)=(0, 0)\cite{SKitagawa_2014,Mukuda_jpsj2014}.
These results indicate that the AFMSFs are one of the important factors for enhancing the $T_c$ in Fe-pnictide SCs, even when the lattice parameters deviate from their optimum values for a Fe$Pn_4$ regular tetrahedron\cite{C.H.Lee,Mizuguchi}. 

On the other hand, the AFMSFs are not so distinct at low energies for the compounds with a high $T_c$($\sim$50 K), which are characterized by the local lattice parameters of FeAs$_4$ close to the regular tetrahedron  \cite{MukudaPRL,Tomita,Yamamoto,Miyamoto}. 
It has been reported that once further electrons are doped in the hydrogen-substituted LaFeAs(O$_{1-y'}$H$_{y'}$), the new phases of SC and AFM orders are uncovered and denoted as SC3 and AFM3 in Fig.~\ref{phasediagram}, respectively \cite{Iimura_H,Hiraishi}. 
The theory has pointed out that the electronic state for the onset of SC3 resembles that of the highest $T_c$($\ge$50 K) state \cite{Suzuki_H}. 
Thus, further systematic studies over wide compositions of $x$ and $y$ in LaFe(As$_{1-x}$P$_x$)(O$_{1-y}$F$_y$) will provide an opportunity to unravel the universal relationship between the presence of AFMSFs and the onset of SC, including the unexpected relationship between the complicated effect of  some electron doping and the deformation of the local structure of Fe$Pn_4$. 

In this Letter, we report on $^{31}$P-NMR studies of LaFe(As$_{1-x}$P$_x$)(O$_{1-y}$F$_{y}$) over  wide compositions for 0$\le x\le$1 and 0$\le y\le$0.14, revealing that AFMSFs are one of the indispensable elements for enhancing $T_c$. 
Systematic measurements of the $^{31}$P nuclear spin relaxation rate $(1/T_1)$ have revealed that the multiple AFMSFs relevant to the multiple-orbital nature of Fe-pnictides are responsible for increasing $T_c$ over wide compositions of $x$ and $y$.
As a result, we remark that a nonmonotonic variation of $T_c$ in  LaFe(As$_{1-x}$P$_x$)(O$_{1-y}$F$_y$) is attributed to the multiple AFMSFs originating from degenerated multiple $3d$ orbitals inherent to Fe-pnictides.


Detailed $^{31}$P-NMR ($I$=1/2) measurements were performed on coarse-powder polycrystalline samples of LaFe(As$_{1-x}$P$_x$)(O$_{0.86}$F$_{0.14}$) with nominal contents at $x$=0.2, 0.4, 0.6, and 1.0. These samples were synthesized by the solid-state reaction method~\cite{Uekubo,Lai_PRB}. 
Powder X-ray diffraction measurements indicated that the lattice parameters exhibit a monotonic variation with $x$\cite{Uekubo}. 
$T_c$s were determined from an onset of SC diamagnetism in the susceptibility measurement, as shown in Fig. \ref{phasediagram}\cite{Uekubo}.  
Extensive studies over a wide composition of LaFe(As$_{1-x}$P$_x$)(O$_{1-y}$F$_{y}$) were also performed on the samples of ($x$, $y$)=(0.2, 0.05)  and  ($x$, $y$)=(0.2, 0.1).
The Knight shift $K$ was measured at a magnetic field of $\sim$11.93 T, which was calibrated using a resonance field of $^{31}$P in H$_3$PO$_4$. 
Generally, $K$ comprises the temperature($T$)-dependent spin shift $K_s(T)$ and the $T$-independent chemical shift $K_{\rm chem}$, expressed as $K=K_s(T)+K_{\rm chem}$. 
The nuclear-spin lattice-relaxation rate $^{31}(1/T_1)$ of $^{31}$P-NMR was measured at the field of $\sim$11.93 T by fitting a recovery curve for $^{31}$P nuclear magnetization to a single exponential function $m(t)\equiv [M_0-M(t)]/M_0=\exp \left(-t/T_1\right)$. Here, $M_0$ and $M(t)$ are the respective  nuclear magnetizations for a thermal equilibrium condition and at time $t$ after a saturation pulse.


\begin{figure}[htbp]
\centering
\includegraphics[width=7cm]{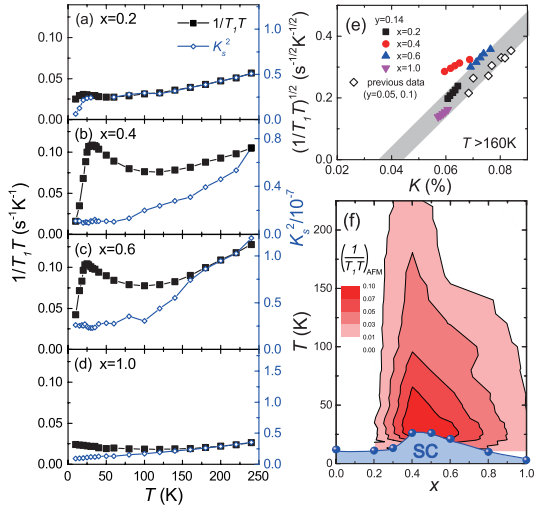}
\caption[]{(Color online) 
$T$ dependences of $(1/T_1T)$ and $K_s^2(T)$ for (a) $x$=0.2, (b) $x$=0.4, (c) $x$=0.6, and (d) $x$=1.0 of LaFe(As$_{1-x}$P$_x$)(O$_{0.86}$F$_{0.14}$). (e) Plot of $(1/T_1T)^{1/2}$ {\it vs} $K$ with an  implicit parameter of $T$ for these compounds, using data at high temperatures ($T\ge$160 K). 
The empty diamonds are the previous data on $y$=0.05 and 0.1 ($T$=200 K).\cite{Mukuda_PRB2014,Mukuda_jpsj2014} 
The data for $x$=0.4 deviate from this line owing to the development of AFMSFs. 
(f) Contour plot of $(1/T_1T)_{\rm AFM}$ for $y$=0.14, indicating that the AFMSFs significantly develop where $T_c$ exhibits a peak. 
}
\label{F14_T1}
\end{figure}

Figures \ref{F14_T1}(a)-\ref{F14_T1}(d) show the $T$ dependences of $(1/T_1T)$ and $K_s^2(T)$($=$($K-K_{\rm chem}$)$^2$) for $x$=0.2, 0.4, 0.6, and 1.0. 
The plot of $(1/T_1T)^{1/2}$ against $K(T)$, as shown in Fig. \ref{F14_T1}(e),  enables us to obtain $K_s$ by evaluating $K_{\rm chem}$ to be 0.04($\pm$0.01)\% for these compounds.
Here, note that the data at temperatures higher than $T\sim$160 K are used since the contribution of AFMSFs in $(1/T_1T)$ is negligible.  
The value of  $K_{\rm chem}$ is comparable to those evaluated in previous studies\cite{Mukuda_PRB2014,Mukuda_jpsj2014,Miyamoto}. 
$K_s(T)$ is proportional to $\chi({q}$=$0)$ with the relation  $K_s(T)=\!A_{\rm hf}(0)\chi({q}$=$0)\propto~\!A_{\rm hf}(0)N(E_{\rm F})$. Here, $\chi({q}$=$0)$ is the static spin susceptibility, $\!A_{\rm hf}$(0) is the hyperfine coupling constant  at ${q}$=0, and $N(E_{\rm F})$ is the density of states (DOS) at the Fermi level ($E_{\rm F}$).  

As for $x$=0.2 and 1.0, the $T$ dependence of $(1/T_1T)$ follows that of $K_s^2(T)$ for a wide $T$ region, as shown in Figs. \ref{F14_T1}(a) and  \ref{F14_T1}(d), which point to the Korringa's relation $(1/T_1T)\propto N(E_{\rm F})^2$ expected for conventional metals. 
By contrast, the $(1/T_1T)$s at $x$=0.4 and 0.6 increase as temperature decreases, although $K_s(T)$s decrease as shown in Figs. \ref{F14_T1}(b) and  \ref{F14_T1}(c). 
This contrasted behavior between  $(1/T_1T)$ and $K_s(T)$ demonstrates the development of AFMSFs with finite wave vectors as temperature decreases. 
To deduce the development of AFMSFs following the previous studies~\cite{Mukuda_PRB2014,Mukuda_jpsj2014,Miyamoto},we assume that $(1/T_1T)$ is decomposed as
\[
(1/T_1T)=(1/T_1T)_{\rm AFM}+(1/T_1T)_{0}, 
\]
where the first term represents the contribution of AFMSFs with the finite wave vectors $Q$ presumably around (0,$\pi$) and ($\pi$,0) that significantly develop upon cooling, and the second one represents the 
$q$-independent one in proportion to $N(E_{\rm F})^2$. 
Figure~\ref{F14_T1}(f) shows a contour plot of $(1/T_1T)_{\rm AFM}$ for $y$=0.14, which is illustrated by assuming that the $T$ dependence of $(1/T_1T)_0$ is identical to that of $K_s^2(T)$.
This contour shows that the AFMSFs develop upon cooling for $x$=0.4 and 0.6 exhibiting relatively high $T_c$ values; by contrast, they are markedly suppressed for $x$=0.2 and 1.0 exhibiting very low $T_c$ values.

\begin{figure*}[tb]
\centering
\includegraphics[width=11cm]{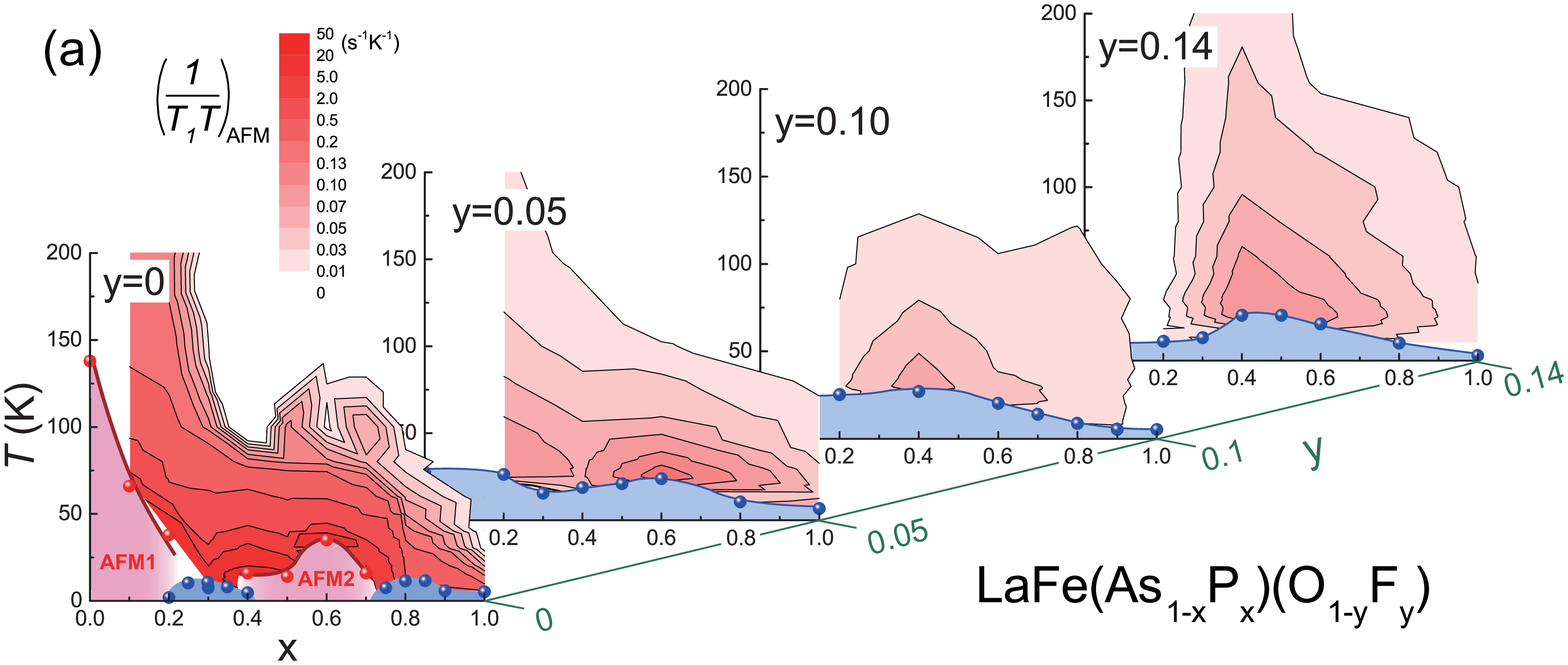}
\includegraphics[width=4cm]{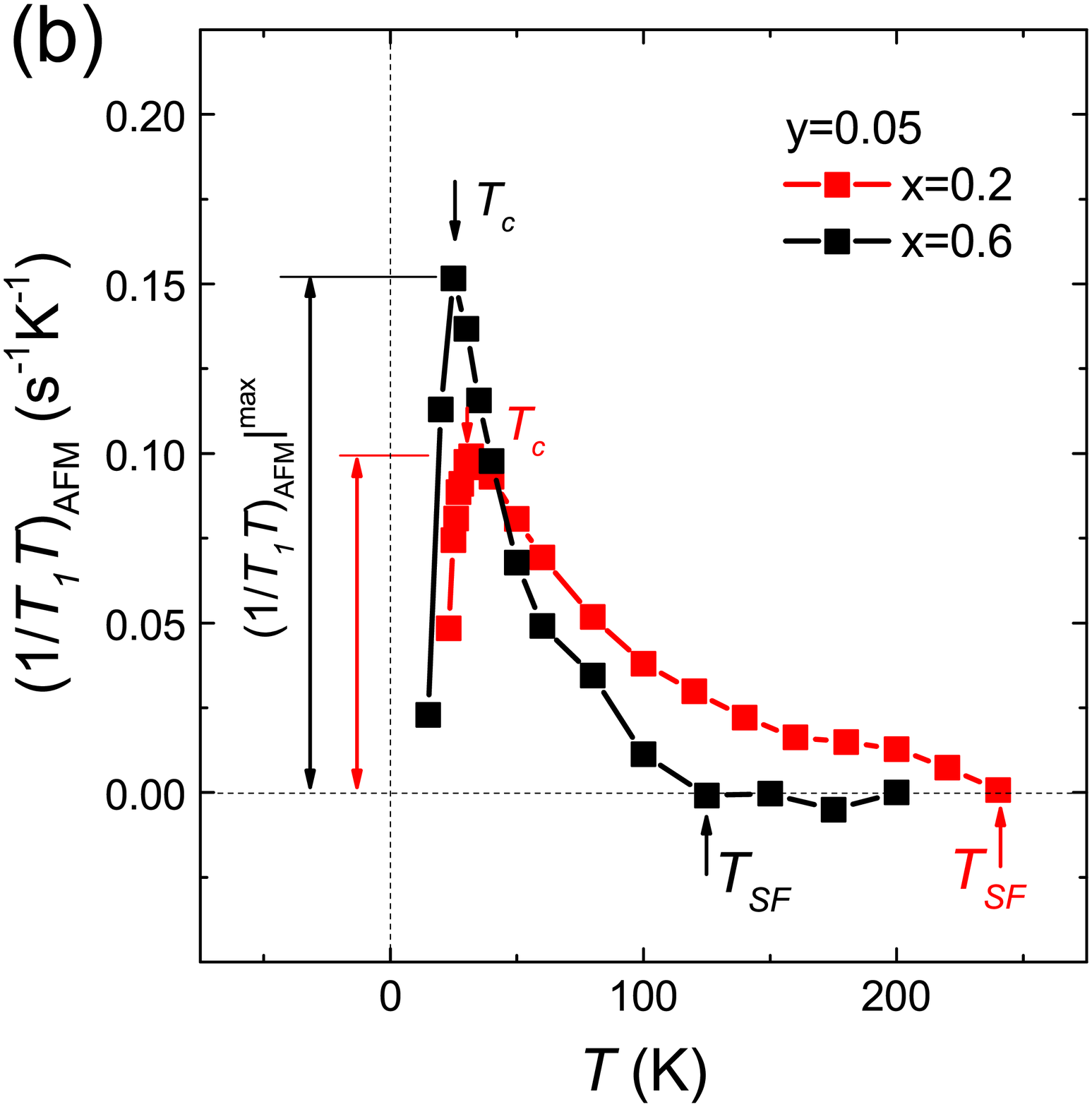}
\caption[]{(Color online) 
(a) Contour plots of $(1/T_1T)_{\rm AFM}$ for $y$=0.0\cite{Mukuda_jpsj2014}, 0.05, 0.10\cite{Mukuda_PRB2014}, and 0.14 in a common scale, providing clear evidence  that the AFMSFs play a significant role in raising $T_c$  for the present compositions of $x$ and $y$. 
(b) Typical $T$ dependence of $(1/T_1T)_{\rm AFM}$ at $x$=0.2 and 0.6 for $y$=0.05. 
The AFMSFs at $x$=0.2  gradually develop upon cooling from higher temperature, whereas the AFMSFs at  $x$=0.6 develop rapidly only at low temperature, although the value of $(1/T_1T)_{\rm AFM}$ at low temperature is smaller in the former case than in the latter case. 
We define $T_{\rm SF}$ as the temperature below which the AFMSFs start to develop, and $(1/T_1T)_{\rm AFM}|^{\rm max}$ as the maximum value of  $(1/T_1T)_{\rm AFM}$.
}
\label{allAFM}
\end{figure*}

Figure~\ref{allAFM}(a) shows the contour plots of $(1/T_1T)_{\rm AFM}$ for 0$\le y \le$0.14 in a common scale.
The AFMSFs play a significant role in raising $T_c$ for wide compositions of $x$ and $y$. 
Here, note that there are two types of AFMSFs in the $T$-evolution upon cooling. 
Namely,  focusing on the $T$-variation of $(1/T_1T)_{\rm AFM}$ for the two samples of $y$=0.05, Fig.~\ref{allAFM}(b) indicates that the AFMSFs at $x$=0.6 largely develop only at low temperatures, whereas those at $x$=0.2 gradually develop upon cooling from high temperatures. 
To gain further insight into these features of the AFMSFs, we define $T_{\rm SF}$ as the temperature below which  AFMSFs start to develop, and $(1/T_1T)_{\rm AFM}|^{\rm max}$ as the maximum value of  $(1/T_1T)_{\rm AFM}$, as presented in Fig.~\ref{allAFM}(b). 
$(1/T_1T)_{\rm AFM}$ dominated by AFMSFs with a wave vector $Q$ is generally described as
\[
\left(\frac{1}{T_1T}\right)_{\rm \!\!AFM} \propto \lim_{\omega\rightarrow \omega_0\sim0}  |A_{\rm hf}(Q)|^2 \frac{\chi''(Q,\omega)}{\omega},
\]
where $A_{\rm hf}(Q)$ is the hyperfine-coupling constant at $q=Q$, $\chi(Q,\omega)$ is the dynamical spin susceptibility at $q=Q$ and energy $\omega$, and $\omega_0$ is an NMR frequency approximating as $\omega_0\sim 0$. 
Thus, the $(1/T_1T)_{\rm AFM}|^{\rm max}$ probes a low energy limit of  $\chi''(Q,\omega)/\omega$ around $\omega_0\sim 0$, representing how large the spectral weight of AFMSFs are at low energies. 
The $T_{\rm SF}$ indicates the temperature that $\chi''(Q,\omega)$ significantly starts to increase upon cooling, roughly pointing to a characteristic energy of AFMSFs. 

\begin{figure*}[thb]
\centering
\includegraphics[width=16cm]{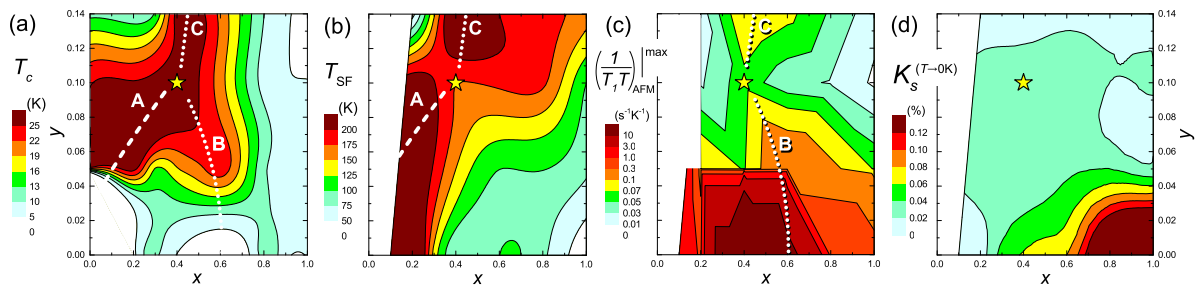}
\caption[]{(Color online) Contour plots of (a) $T_c$~\cite{Kamihara2008,Hosono_review,Luetkens,Saijo,Lai_PRB,Wang,Uekubo}, (b) $T_{\rm SF}$, (c) $(1/T_1T)_{\rm AFM}|^{\rm max}$, and (d) $K_s$($T\!\!\rightarrow$0)  for wide compositions of $x$ and $y$ in  LaFe(As$_{1-x}$P$_x$)(O$_{1-y}$F$_{y}$).
The definitions of $T_{\rm SF}$ and $(1/T_1T)_{\rm AFM}|^{\rm max}$ are shown in Fig. \ref{allAFM}(b).   
The respective curves  A and B correspond to the branch that $T_{\rm SF}$ is relatively high and the one that $(1/T_1T)_{\rm AFM}|^{\rm max}$ is  relatively large.
The star  at  ($x$, $y$)=(0.4, 0.1) indicates the highest $T_c$(=27 K) point in the present compounds. 
}
\label{all}
\end{figure*} 

Note that  the curves  A, B,  and C in Fig.~\ref{all}(a) are along the values of $T_c$ being relatively high for the wide compositions of $x$ and $y$\cite{Kamihara2008,Hosono_review,Luetkens,Saijo,Lai_PRB,Wang,Uekubo}. 
Figures \ref{all}(b) and \ref{all}(c) show the contour plots of $T_{\rm SF}$ and $(1/T_1T)_{\rm AFM}|^{\rm max}$, respectively. 
The curve A  is along the high values of $T_{\rm SF}$  in Fig.~\ref{all}(b), demonstrating that the development of AFMSFs below $T_{\rm SF}$ is mostly responsible for increasing $T_c$. 
Here, note that the high values of $T_c$ are along the high values of $T_{\rm SF}$, in other words, roughly along  the high values of the characteristic energy of AFMSFs. 
In fact, the theory predicted that the three orbitals of $d_{xz\!/yz}$+$d_{xy}$ are relevant to the AFMSFs characteristic at finite energies rather than at low energies in association with the AFM1 order at ($x$, $y$)=(0, 0) \cite{Kuroki2,Usui_SR,Arai}.
On the other hand, the high values of $T_c$ along the curve B  are along the large values of $(1/T_1T)_{\rm AFM}|^{\rm max}$, as shown in Fig.~\ref{all}(c), demonstrating that the development of the AFMSFs at low energies is also responsible for increasing $T_c$, originating from the collapse of the AFM2 order at ($x$, $y$)=(0.6, 0)\cite{Mukuda_jpsj2014,SKitagawa_2014}. 
Here, note that the very good nesting of the hole Fermi surfaces (FSs) at $\Gamma$(0,0) and electron FSs at $M$(0,$\pi$)($\pi$,0) in the unfolded FS regime is dominated mostly by the two orbitals of $d_{xz\!/yz}$, bringing about the onset of the AFM2 order at ($x$, $y$)=(0.6, 0)\cite{Usui_SR,Arai}.  
Hence, the present result suggests that these AFMSFs dominated by the two  orbitals of $d_{xz\!/yz}$ are gradually suppressed against the increase in the electron doping level as $y$ increases from 0 to 0.14.

Theoretically, two different types of the $T$-evolution of AFMSFs shown in Fig.~\ref{allAFM}(b) were consistently reproduced by the fluctuation-exchange (FLEX) approximation in the multi-orbital Hubbard model. 
In this model, Arai {\it et al.} revealed that the $d_{xz\!/yz}$-derived AFMSFs around the AFM2 phase are largely enhanced at low energies, whereas the $d_{xz\!/yz}$+$d_{xy}$ derived AFMSFs around the AFM1 are characteristic at finite energies rather than at low energies\cite{Arai}. 
In this context, it is notable that the highest $T_c$=27 K is denoted at ($x$, $y$)=(0.4, 0.1) by a star in Fig. \ref{all}(a) around which the curves A and B merge. 
It is instructive to note that the increase in the characteristic energy of AFMSFs from low to high energies brings about the increase in $T_c$. 
This event is consistent with the spin-fluctuations mediated SC mechanism, which enables us to calculate a possible value of $T_c$ on the basis of the integration of the AFMSF spectrum over a wide energy range. 

We also note that as seen in Fig.~\ref{all}(a),  the high value of $T_c$ at ($x$, $y$)=(0.4, 0.1) is kept along the curve C toward ($x$, $y$)=(0.4, 0.14) at which the AFMSFs from low to high energies slightly recover, as deduced from  Figs.~\ref{all}(b) and \ref{all}(c).  
This event  may be associated with a reemergence of the $d_{xz\!/yz}$-derived AFMSFs since the low pnictogen height at large $x$($\sim$0.5) will cause  the FSs  of the $d_{xy}$ orbital to sink below $E_{\rm F}$ and the nesting of FSs of $d_{xz\!/yz}$ orbitals to become somewhat better\cite{Usui_SR}.  
It is worth comparing this with the further electron-overdoped SC3 state in LaFeAs (O$_{1-y'}$H$_{y'}$)\cite{Iimura_H} shown in Fig. \ref{phasediagram}, since no nesting of FSs is expected.
In these compounds, note that the hole FS in association with the $d_{xz\!/yz}$  orbitals significantly shrinks owing to the heavy electron doping, whereas there still remain the hole FS relevant to the $d_{xy}$  orbital  and the large electron FSs\cite{Iimura_H}. 
According to the spin-fluctuation model\cite{Suzuki_H}, the prioritized diagonal hopping on the $d_{xy}$ orbitals  reenhances the other type of AFMSFs in the high-$T_c$ state of the SC3 phase, which is dominated by the high-energy AFMSFs\cite{Iimura_AF}, rather than the low-energy ones \cite{Sakurai}. 
It is interesting to unravel the systematic relationship between $T_c$ and the evolution of different types of AFMSFs in the series going from LaFe(As$_{1-x}$P$_x$)(O$_{1-y}$F$_{y}$) to LaFeAs(O$_{1-y'}$H$_{y'}$). 

Finally, we remark on the variation of $N(E_{\rm F})$ over wide $x$ and $y$ regions, which can be seen from the contour plot of $K_s$($T\!\!\!\rightarrow$0)  estimated from an extrapolation to $T\!\!\!\rightarrow$0, as shown in Fig.~\ref{all}(d). 
Note that $K_s$($T\!\!\!\rightarrow$0)  is directly proportional to $\chi(q$=$0)$ or $N(E_{\rm F})$. 
As seen in the figure, $K_s$($T\!\!\!\rightarrow$0)  increases around ($x$, $y$)=(1, 0) owing to the peak of DOS mainly arising from the $d_{3z^2-r^2}$-derived three-dimensional hole pocket around Z($\pi$,$\pi$,$\pi$)\cite{Miyake,Mukuda_jpsj2014}.
In fact, this contour plot of $N(E_{\rm F})$ has no correlation with  the $T_c$ values.
This indicates that the BCS-type SC mechanism through the electron-phonon interaction is not applicable even for the phosphorus-end members ($x$=1.0), where the AFMSFs are significantly reduced and their $T_c$ values  are less than 10K.  

In summary, the systematic $^{31}$P-NMR measurements for LaFe(As$_{1-x}$P$_x$) (O$_{1-y}$F$_{y}$) with 0$\le x \le$1 and 0$\le y \le$0.14  have unraveled two types of AFMSFs in the $T$-evolution upon cooling, that is, one is the AFMSFs that develops rapidly down to $T_c$ with low-energy characteristics, and  the other, with relatively higher energy than the former, develops gradually upon cooling from high temperature. 
The low-energy AFMSFs in low $y$ (electron doping) over a wide $x$ (pnictogen-height suppression) range are associated with the nesting effect of FSs dominated mostly by the two orbitals of $d_{xz\!/yz}$, whereas the higher-energy ones for a wide $y$ region around low $x$ originate from the three orbitals of $d_{xy}$ and $d_{xz\!/yz}$ \cite{Usui_SR,Arai}. 
The intimate correlation between multiple AFMSFs and $T_c$ values indicates that the AFMSFs are one of the indispensable elements for enhancing $T_c$, even though the lattice parameters deviate from their optimum values for the Fe$Pn_4$ regular tetrahedron.
We remark that the nonmonotonic variation of $T_c$ as a function of $x$ and $y$ in  LaFe(As$_{1-x}$P$_x$)(O$_{1-y}$F$_y$) is attributed to these multiple AFMSFs originating from degenerated multiple $3d$ orbitals inherent to Fe-pnictide superconductors.


{\footnotesize 
This work was supported by Grants-in-Aid for Scientific Research (Nos. 26400356 and 26610102) from the Ministry of Education, Culture, Sports, Science and Technology (MEXT), Japan. }


\end{document}